\documentclass{svjour3}
\smartqed  
\usepackage{graphicx}
\usepackage{amsmath}
\usepackage{amsfonts}
\usepackage{amssymb}
\newcommand{\beq}{\begin{equation}}
\newcommand{\eeq}{\end{equation}}
\newcommand{\beqa}{\begin{eqnarray}}
\newcommand{\eeqa}{\end{eqnarray}}
\journalname{Few-Body Systems (FB20)}
\begin{document}

\title{Recent Progress in the Theory of Nuclear Forces
\thanks{Presented at the 20th International IUPAP Conference on Few-Body Problems in Physics, 20 - 25 August, 2012, Fukuoka, Japan}
}


\author{R. Machleidt
        \and
        Q. MacPherson
           \and
           E. Marji
              \and
              R. Winzer
                 \and
                 Ch. Zeoli
                 \and
                 D. R. Entem
}


\institute{R. Machleidt   \and Q. MacPherson \and E. Marji  \and R. Winzer \and Ch. Zeoli
              \at Department of Physics, University of Idaho, Moscow, ID 83844-0903, USA \\
              \email{machleid@uidaho.edu}           
              \and
           \emph{Present address:} Ch. Zeoli
           \at  Department of Physics, Florida State University, Tallahassee, FL 32306-4350, USA
          \and
           \emph{Present address:} E. Marji
          \at College of Western Idaho, Nampa, ID 83653, USA
          \and
           D. R. Entem \at
               Grupo de Fisica Nuclear and IUFFyM, University of Salamanca, E-37008 Salamanca, Spain \\
                \email{entem@usal.es}  
}

\date{Received: date / Accepted: date}

\maketitle

\begin{abstract}
During the past two decades, it has been demonstrated that chiral effective field theory represents a powerful tool to deal with nuclear forces in a systematic and model-independent way. Two-, three-, and four-nucleon forces have been derived up to next-to-next-to-next-to-leading order (N$^3$LO)
and (partially) applied in nuclear few- and many-body
systems---with, in general, a good deal of success. This may suggest that we are finally done with the nuclear
force problem; but that would be too optimistic. There are still some pretty basic open issues
that have been swept under rug and, finally, need our full attention, like the proper renormalization of the
two-nucleon potential. Moreover, the order-by-order convergence of the many-body force contributions is at best obscure at this time.
\keywords{Effective field theory \and Chiral perturbation theory \and Renormalization \and 
Nuclear two- and many-body forces}
\end{abstract}

\section{Introduction}

The theory of nuclear forces has a long history (cf.\ Table~\ref{tab_hist}).
Based upon the seminal idea by Yukawa~\cite{Yuk35}, 
first field-theoretic attempts
to derive the nucleon-nucleon ($NN$) interaction
focused on pion exchange.
While the one-pion exchange turned out to be very useful
in explaining $NN$ scattering data and, in particular, the properties
of the deuteron~\cite{Sup56},
multi-pion exchange was beset with
serious ambiguities~\cite{TMO52}
that could not be resolved in a satisfactory way.
Thus, the ``pion theories'' of the 1950s
are generally judged as failures---for reasons
we understand today: pion dynamics is constrained by chiral
symmetry, a crucial point that was unknown in the 1950s.

\begin{table}[t]
\caption{The Theory of Nuclear Forces:
Eight Decades of Struggle
\label{tab_hist}} 
\smallskip
\begin{tabular*}{\textwidth}{@{\extracolsep{\fill}}cc}
\hline 
\hline 
\noalign{\smallskip}
\bf 1935   &
\bf Yukawa: Meson Theory
\\
\noalign{\smallskip}
\hline
\noalign{\smallskip}
      &            
{\it The ``Pion Theories''.}
\\
      \bf 1950's     &
One-pion exchange: good;\\
   & 
Multi-pion exchange: disaster.
\\
\noalign{\smallskip}
\hline
\noalign{\smallskip}
\bf 1960's & 
Many pions $\equiv$ multi-pion resonances:
{\boldmath $\sigma$, $\rho$, $\omega$, ...}
\\
           & 
The One-Boson-Exchange Model: success.
\\
\noalign{\smallskip}
\hline
\noalign{\smallskip}
\bf 1970's & 
            Refinement of meson theory:
Sophisticated {\boldmath $2\pi$} exchange models;
\\
           & 
Partovi-Lomon, Stony Brook, Paris, Bonn.
\\
\noalign{\smallskip}
\hline
\noalign{\smallskip}
\bf 1980's &  
Nuclear physicists discover {\bf QCD}:
\\
           & Quark Models.
\\
\noalign{\smallskip}
\hline
\noalign{\smallskip}
\bf 1990's & 
Nuclear physicists discover {\bf EFT}; Weinberg, van Kolck, \ldots
\\
\bf and beyond &
{\bf Back to Yukawa's Meson (Pion) Theory!} \\
               & {\it But, constrained by Chiral Symmetry.}
\\
\noalign{\smallskip}
\hline
\hline
\end{tabular*}
\vspace*{0.5cm}
\end{table}

Historically, the experimental discovery of heavy 
mesons in the early 1960s
saved the situation. The one-boson-exchange (OBE)
model emerged, which still today is the most economical
and quantitative
phenomenology for describing the 
$NN$ interaction~\cite{Mac89}.
The weak point of this model, however, is the scalar-isoscalar
``sigma'' or ``epsilon'' boson, for which empirical
evidence remains controversial. Since this boson is associated
with the  correlated (or resonant) exchange of two pions,
a vast theoretical effort 
was launched to derive the 2$\pi$-exchange contribution
of the nuclear force, which creates the intermediate 
range attraction.
During this effort, which occupied more than a decade,
dispersion theory 
(Stony Brook and Paris potentials) 
as well as field theory 
(Partovi-Lomon model, Bonn potential~\cite{Mac89,MHE87})
were invoked.

The nuclear force problem appeared to be solved; however,
with the discovery of quantum chromo-dynamics (QCD), 
all ``meson theories'' were
relegated to the status of models and the attempts to derive
the nuclear force had to start all over again.

The problem with a derivation of nuclear forces from QCD is that
this theory is non-perturbative in the low-energy regime
characteristic of nuclear physics, which makes direct solutions
very difficult.
Therefore, during the first round of new attempts, QCD-inspired quark 
models
became popular. 
The positive aspect of these models is that they try to explain hadron structure
and hadron-hadron interactions on an equal footing and, indeed, 
some of the gross features of the $NN$ interaction are explained successfully.
However, on a critical note, it must be pointed out
that these quark-based
approaches are nothing but
another set of models and, thus, do not represent
fundamental progress. 
For the purpose of describing hadron-hadron interactions, 
one may equally well stay
with the simpler and much more quantitative meson models.

A major breakthrough occurred when 
the concept of an effective field theory (EFT) was introduced
and applied to low-energy QCD.
As outlined by Weinberg in a seminal paper~\cite{Wei79},
one has to write down the most general Lagrangian consistent
with the assumed symmetry principles, particularly
the (broken) chiral symmetry of QCD. At low energy, the effective degrees of freedom are pions 
(the Goldstone bosons of the broken symmetry) and
nucleons rather than quarks and gluons; heavy mesons and
nucleon resonances are ``integrated out''.
So, the circle of history is closing and we are {\it back to Yukawa's meson (pion) theory},
except that we have finally learned how to deal with it:
broken chiral symmetry is a crucial constraint that generates
and controls the dynamics and establishes a clear connection
with the underlying theory, QCD.

The past 15 years have seen great progress in applying chiral perturbation theory (ChPT)
to nuclear forces
\cite{Wei90,Wei92,ORK94,Kol94,KBW97,EM03,EGM05,ME11}.
As a result, nucleon-nucleon ($NN$) potentials of high precision have been constructed, which
are based upon ChPT carried to next-to-next-to-next-to-leading order 
(N$^3$LO)~\cite{EM03,EGM05,ME11}, and applied in nuclear structure calculations
with great success.

However, in spite of this progress, we are not done. Due to the complexity of the
nuclear force issue, there are still many subtle and not so subtle open problems.
We will not list and discuss all of them, but instead just focus on the two open issues, 
which we perceive as the most important ones:
\begin{itemize}
\item
The proper renormalization of chiral nuclear potentials and
\item
Subleading chiral few-nucleon forces.
\end{itemize}

\section{Renormalization of chiral nuclear forces}

\subsection{The chiral $NN$ potential}
In terms of naive dimensional analysis or ``Weinberg counting'',
the various orders of the irreducible graphs which define the chiral $NN$ potential 
are given by:
\beqa
V_{\rm LO} & = & 
V_{\rm ct}^{(0)} + 
V_{1\pi}^{(0)} 
\label{eq_VLO}
\\
V_{\rm NLO} & = & V_{\rm LO} +
V_{\rm ct}^{(2)} + 
V_{1\pi}^{(2)} +
V_{2\pi}^{(2)} 
\label{eq_VNLO}
\\
V_{\rm NNLO} & = & V_{\rm NLO} +
V_{1\pi}^{(3)} + 
V_{2\pi}^{(3)} 
\label{eq_VNNLO}
\\
V_{{\rm N}^3{\rm LO}} & = & V_{\rm NNLO} +
V_{\rm ct}^{(4)} +
V_{1\pi}^{(4)} +  
V_{2\pi}^{(4)} +
V_{3\pi}^{(4)} 
\label{eq_VN3LO}
\eeqa
where 
the superscript denotes the order $\nu$ of the low-momentum
expansion.
LO stands for leading order, NLO for next-to-leading
order, etc..
Contact potentials carry the subscript ``ct'' and
pion-exchange potentials can be identified by an
obvious subscript. For more details concerning the above
potentials, see ref.~\cite{ME11}.

\subsection{Nonperturbative renormalization of the $NN$ potential}
The two-nucleon system is characterized by large scattering lengths and shallow (quasi)
bound states which require a nonperturbative treatment.
Following Weinberg's prescription~\cite{Wei90}, this is accomplished by
inserting the potential $V$ into the Lippmann-Schwinger (LS) equation:
\begin{equation}
 {T}({\vec p}~',{\vec p})= {V}({\vec p}~',{\vec p})+
\int d^3p''\:
{V}({\vec p}~',{\vec p}~'')\:
\frac{M_N}
{{ p}^{2}-{p''}^{2}+i\epsilon}\:
{T}({\vec p}~'',{\vec p}) \,,
\label{eq_LS}
\end{equation}
where $M_N$ denotes the nucleon mass.

In general, the integral in
the LS equation is divergent and needs to be regularized.
One way to achieve this is  by
multiplying $V$
with a regulator function
\begin{equation}
{ V}(\vec{ p}~',{\vec p}) 
\longmapsto
{ V}(\vec{ p}~',{\vec p})
\;\mbox{\boldmath $e$}^{-(p'/\Lambda)^{2n}}
\;\mbox{\boldmath $e$}^{-(p/\Lambda)^{2n}}
\label{eq_regulator} \,.
\end{equation}
Typical choices for the cutoff parameter $\Lambda$ that
appears in the regulator are 
$\Lambda \approx 0.5 \mbox{ GeV} < \Lambda_\chi \approx 1$ GeV.

In field theories, divergent integrals are not uncommon and methods have
been designed to deal with them.
One regulates the integrals and then removes the dependence
on the regularization parameters (scales, cutoffs)
by ``renormalization''. In the end, the theory and its
predictions do not depend on cutoffs
or renormalization scales.
So-called renormalizable quantum field theories, like QED,
have essentially one set of prescriptions 
that takes care of renormalization through all orders. 
In contrast, 
EFTs are renormalized by ``counter terms'' (contact terms) that are introduced order by order
in increasing numbers.

Naively, the most perfect renormalization procedure is the one where the cutoff
parameter $\Lambda$ is taken to infinity while stable and quantitative results are maintained
through the adjustment of counter terms.
This was accomplished at LO in the work by Nogga {\it et al}~\cite{NTK05}.
At NNLO, the infinite-cutoff renormalization procedure has been investigated 
in~\cite{YEP07} for partial waves with total angular momentum $J\leq 1$ and
in~\cite{VA07} for all partial waves with $J\leq 5$. 
However, for a quantitative chiral $NN$ potential one needs to advance all the way
to N$^3$LO.
At N$^3$LO,
the $^1S_0$ state was considered in Ref.~\cite{Ent08},
and all states up to $J=6$ were investigated in Ref.~\cite{ZME12}.
From all of these works, it is evident that no counter term is effective in partial-waves with
short-range repulsion and only a single counter term can constructively be used in
partial-waves with short-range attraction. Thus, for the $\Lambda \rightarrow \infty$
renormalization prescription, even at N$^3$LO, there exists either one or no counter term
per partial-wave state. This is inconsistent with any reasonable power-counting scheme
and prevents an order-by-order improvement of the predictions.

To summarize:
In the infinite-cutoff renormalization scheme, the potential is admitted up to unlimited momenta.  However, the EFT this potential is derived from has validity only for momenta smaller than the chiral symmetry breaking scale $\Lambda_{\chi}\approx$ GeV.  The lack of order-by-order convergence and discrepancies in lower partial-waves demonstrate that the potential should not be used beyond the limits of the effective theory~\cite{ZME12} (see Ref.~\cite{EG09}
for a related discussion).  The conclusion then is that cutoffs should be limited to $\Lambda\lesssim\Lambda_{\chi}$
(but see also Ref.~\cite{EG12} and J. Gegelia's contribution to this conference).

Crucial for an EFT are regulator independence (within the range of validity
of the EFT) and a power counting scheme that allows for order-by-order
improvement with decreasing truncation error.
The purpose of renormalization is to achieve this regulator independence while maintaining
a functional power counting scheme.

\begin{figure*}
\vspace*{-2.7cm}
\includegraphics[scale=.5]{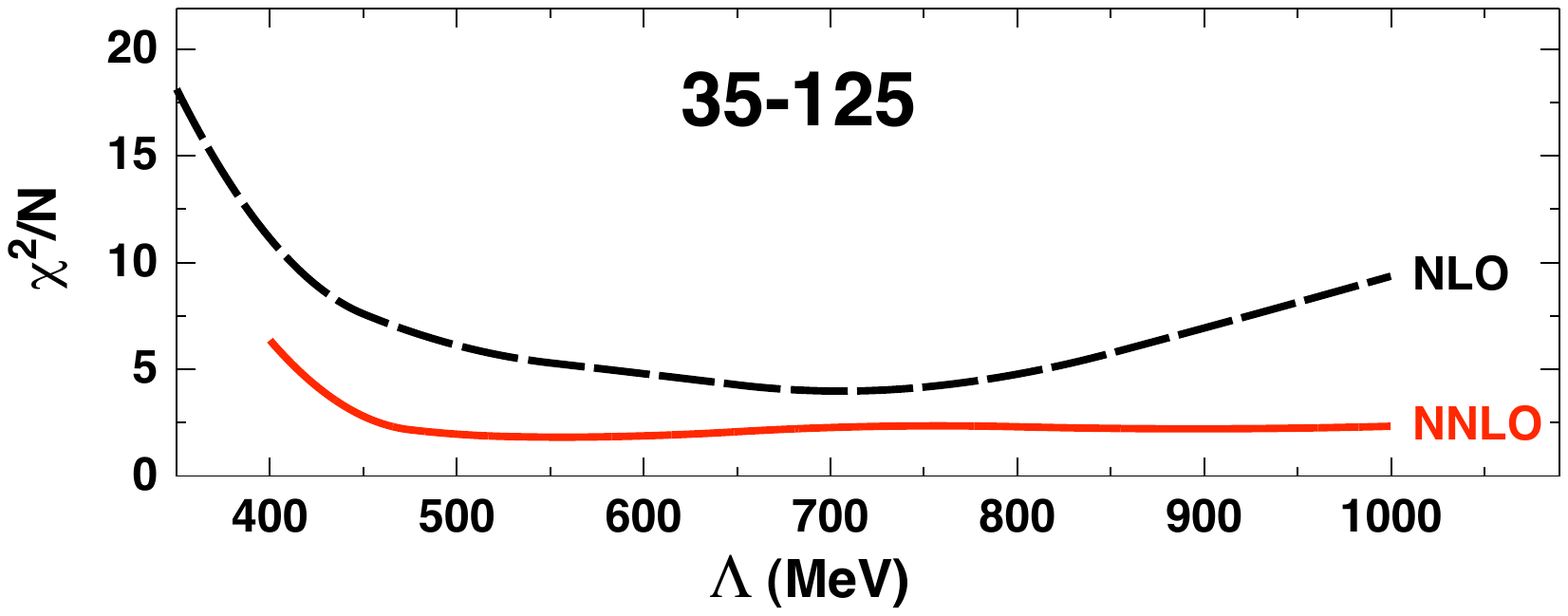}

\vspace*{-9.5cm}
\includegraphics[scale=.5]{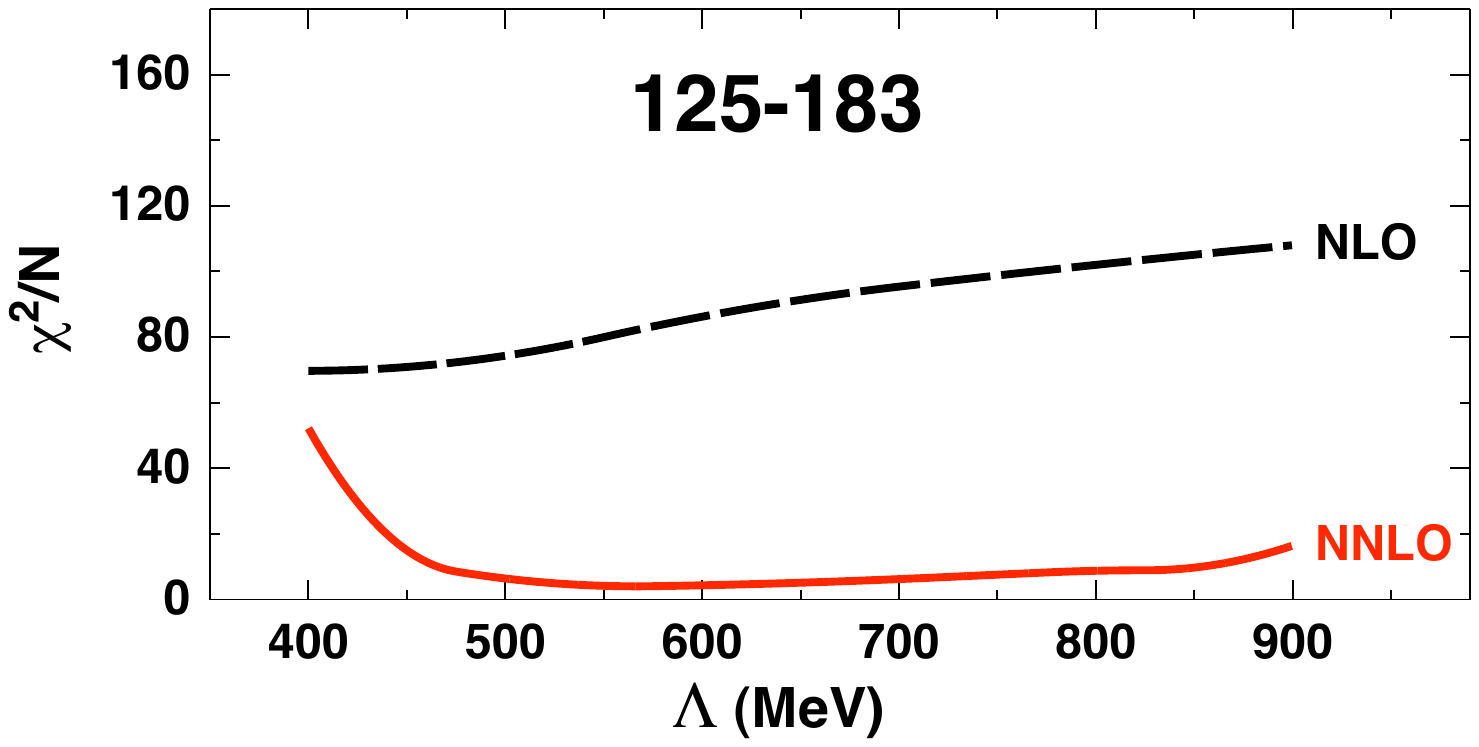}
\vspace*{-7.5cm}
\caption{$\chi^2$/datum for the reproduction of the $np$ data in the
energy range 35-125 MeV (upper frame) and 125-183 MeV (lower frame)
as a function of the cutoff parameter $\Lambda$ of the regulator function
Eq.~(\ref{eq_regulator}). The (black) dashed curves show the $\chi^2$/datum
achieved with $np$ potentials constructed at order NLO and the (red) solid curves are for NNLO.}
\label{fig_32}       
\end{figure*}

Thus, in the spirit of Lepage~\cite{Lep97}, the cutoff independence should be examined
for cutoffs below the hard scale and not beyond. Ranges of cutoff independence within the
theoretical error are to be identified using `Lepage plots'~\cite{Lep97}.
Recently, we have started a systematic investigation of this kind.
In our work, we quantify the error of the predictions by calculating the $\chi^2$/datum 
for the reproduction of the neutron-proton ($np$) elastic scattering data
as a function of the cutoff parameter $\Lambda$ of the regulator function
Eq.~(\ref{eq_regulator}). We have investigated the predictions by chiral $np$ potentials at 
order NLO and NNLO applying Weinberg counting for the counter terms ($NN$ contact terms).
We show our results for the energy range 35-125 MeV in the upper frame of Fig.~\ref{fig_32}
and for 125-183 MeV in the lower frame. 
It is seen that the reproduction of the $np$ data at these energies is generally poor
at NLO, while at NNLO the $\chi^2$/datum assumes acceptable values (a clear demonstration of
order-by-order improvement). Moreover, at NNLO one observes ``plateaus'' of constant low $\chi^2$ for
cutoff parameters ranging from 450 to 850 MeV. This may be perceived as cutoff independence
(and, thus, successful renormalization) in the relevant range of cutoff parameters.

\section{Few-nucleon forces and what is missing \label{sec_manyNF}}

We will now discuss the other issue we perceive as unfinished and important, namely,
subleading chiral few-nucleon forces.

Nuclear three-body forces in ChPT were initially discussed
by Weinberg~\cite{Wei92}.
The three-nucleon force (3NF) at NNLO, was derived by van Kolck~\cite{Kol94}
and applied, for the first time, in nucleon-deuteron
scattering by Epelbaum {\it et al.}~\cite{Epe02b}.
The leading 4NF (at N$^3$LO) was constructed by
Epelbaum~\cite{Epe06} and found to contribute in the
order of 0.1 MeV to the $^4$He binding energy
(total $^4$He binding energy: 28.3 MeV)
in a preliminary calculation~\cite{Roz06}, confirming the traditional
assumption that 4NF are essentially negligible.
{\bf Therefore, the focus is on 3NF.}

For the order of a 3NF, we have
\begin{equation}
\nu = 2 + 2L + 
\sum_i \Delta_i \,,
\label{eq_nu3nf}
\end{equation}
where $L$ denotes the number of loops and $\Delta_i$
is the vertex index.
We will use this equation to analyze 3NF contributions
order by order.
The first non-vanishing 3NF occurs at $\nu=3$ (NNLO), which
is obtained when
there are no loops ($L=0$) and 
$\sum_i \Delta_i = 1$, i.e., 
$\Delta_i=1$ for one vertex 
while $\Delta_i=0$ for all other vertices.
There are three topologies which fulfill this condition,
known as the two-pion exchange (2PE), one-pion exchange (1PE),
and contact graphs.

The 3NF at NNLO
has been applied in
calculations of few-nucleon reactions~\cite{Kal12},
structure of light- and medium-mass 
nuclei~\cite{Nav07,Ots09,HKW09,Rot12,Hag12a,Hag12b},
and nuclear and neutron matter~\cite{HS10,Heb11,Sam12,Cor12}
with a great deal of success.
However, the famous `$A_y$ puzzle' of nucleon-deuteron scattering~\cite{Epe02b} and 
the analogous problem with the
analyzing power in $p$-$^3$He scattering~\cite{Viv10} 
is not resolved.
Furthermore, the spectra of light nuclei leave room for improvement~\cite{Nav07}.
Since we are dealing with a perturbation theory, it is natural to turn to the next
order when looking for improvements.

The next order is N$^3$LO, where we have loop and tree diagrams.
For the loops, we have
$L=1$ and, therefore, all $\Delta_i$ have to be zero
to ensure $\nu=4$. 
Thus, these one-loop 3NF diagrams can include
only leading order vertices, the parameters of which
are fixed from $\pi N$ and $NN$ analysis.
One sub-group of these diagrams (the 2PE graphs)
has been calculated by Ishikawa and Robilotta~\cite{IR07},
and the other topologies
have been evaluated by the Bochum-Bonn group~\cite{Ber08,Ber11}.
The N$^3$LO 2PE 3NF has been applied in the calculation
of nucleon-deuteron observables in Ref.~\cite{IR07} 
causing little impact.
Very recently, the long-range part of the chiral N$^3$LO 3NF has been
tested in the triton~\cite{Ski11} and in three-nucleon scattering~\cite{Wit12}
yielding only moderate effects. The long- and short-range parts of this
force have been used in neutron matter calculations
(together with the N$^3$LO 4NF) producing relatively large contributions
from the 3NF~\cite{Tew12}. Thus, the ultimate assessment of the N$^3$LO 3NF is still
outstanding and will require more few- and many-body applications.

In the meantime, it is of interest to take already a look
at the next order of 3NFs, which is N$^4$LO or $\nu=5$
(of the $\Delta$-less theory to which  the present discussion
is restricted because of lack of space). 
The loop contributions that occur at this order
are obtained by replacing in the N$^3$LO loops
one vertex by a $\Delta_i=1$ vertex (with LEC $c_i$),
which is why these loops may be more sizable than the N$^3$LO loops.
The 2PE topology has already been evaluated~\cite{KGE12} and turns out to be of modest size;
moreover, it can be handled in a practical way by summing it up
together with the 2PE topologies at NNLO and N$^3$LO~\cite{KGE12}.
However, there are four more loop topologies, which are very involved and that have 
not been worked out yet. Finally, a tree topology at N$^4$LO provides
a new set of 3N contact interactions, which have recently been derived
by the Pisa group~\cite{GKV11}. Contact terms are typically simple (as compared
to loop diagrams) and their coefficients are unconstrained (except for naturalness). 
{\it Therefore, it would be an
attractive project to test some terms (in particular, the spin-orbit terms) 
of the N$^4$LO contact 3NF~\cite{GKV11} in calculations of few-body reactions (specifically,
the p-d and p-$^3$He $A_y$) and spectra of light nuclei.}

\section{Conclusions and Outlook}

The past 15 years have seen great progress in our understanding of nuclear forces
in terms of low-energy QCD. Key to this development was the realization that
low-energy QCD is equivalent to an effective field theory (EFT) which allows for 
a perturbative expansion that has become known as chiral perturbation theory (ChPT).
In this framework, two- and many-body forces emerge on an equal footing and the empirical fact
that nuclear many-body forces are substantially weaker then the two-nucleon force
is explained automatically.

In spite of the great progress and success of the past 15 years, there are still some
unresolved issues. One problem is the
proper renormalization of the chiral two- and many-nucleon potentials, where systematic
investigations are already under way (cf.\ Sec.~2).

The other unfinished business are the few-nucleon forces beyond NNLO (``sub-leading
few-nucleon forces'') which are needed to hopefully resolve some important outstanding
nuclear structure problems. At orders N$^3$LO and N$^4$LO very many
new 3NF structures appear, some of which have already been tested.
However, in view of the multitude of 3NF topologies it will take a while until
we will have a proper overview of impact and convergence of these contributions.

If the open issues discussed in this paper will be resolved within
the next few years, then, after 80 years of desperate struggle, we
may finally claim that the nuclear force problem is essentially under control.
The greatest beneficiaries of such progress will be the fields of 
exact few-nucleon calculations and {\it ab initio} nuclear
structure physics.

\begin{acknowledgements}
This work was supported in part by the U.S. Department of Energy
under Grant No.~DE-FG02-03ER41270.
The work of D. R. E. was funded by the Ministerio de Ciencia y
Tecnolog\'\i a under Contract No.~FPA2007-65748, the Junta de Castilla
y Le\'on under Contract No.~GR12,  and
the European Community-Research Infrastructure Integrating
Activity ``Study of Strongly Interacting Matter'' (HadronPhysics2
Grant No.~227431).
\end{acknowledgements}


\end{document}